\documentclass[english,english,aps,pre,superscriptaddress,twocolumn,longtable]{revtex4}
\usepackage[T1]{fontenc}
\usepackage{graphicx}
\usepackage{amssymb}

\makeatletter

\providecommand{\tabularnewline}{\\}

\usepackage{graphicx}
\usepackage{ae} 
\usepackage{epsfig}
\usepackage{graphicx}
\usepackage{dcolumn}
\usepackage{bm}

\usepackage{babel}
\makeatother

\begin{document}

\title{Orbital-Free Density Functional Theory: Linear Scaling Methods \\ for Kinetic Potentials, and Applications to Solid Al and Si}

\author{Jeng-Da Chai\footnote{E-mail: jdchai@berkeley.edu. Present address: Department of Chemistry, 
University of California at Berkeley, and Chemical Sciences Division, 
Lawrence Berkeley National Laboratory, Berkeley, California 94720}}
\affiliation{Institute for Physical Science and Technology, University of Maryland, College Park, Maryland 20742}

\author{Vincent L. Lign\`eres}
\author{Gregory Ho}
\affiliation{Department of Chemistry, Princeton University, Princeton, New Jersey 08544}

\author{Emily A. Carter}
\affiliation{Department of Mechanical and Aerospace Engineering and Program in Applied 
and Computational Mathematics, Princeton University, Princeton, New Jersey 08544}

\author{John D. Weeks}
\affiliation{Institute for Physical Science and Technology, University of Maryland,
College Park, Maryland 20742}
\affiliation{Department of Chemistry and Biochemistry, University of Maryland, 
College Park, Maryland 20742}

\date{\today{}}

\begin{abstract}
In orbital-free density functional theory the kinetic potential (KP), the
functional derivative of the kinetic energy density functional, appears in the
Euler equation for the electron density and may be more amenable to simple
approximations. We study properties of two solid-state systems, Al and
Si, using two nonlocal KPs that gave good results for atoms. Very accurate
results are found for Al, but results for Si are much less satisfactory,
illustrating the general need for a better treatment of extended covalent
systems. A different integration pathway in the KP formalism may prove useful
in attacking this fundamental problem.
\end{abstract}
\maketitle
\newpage

\section{Introduction}

Density-functional theory (DFT) is one of the most useful methods for studying 
ground state properties of many-electron systems. In principle, as shown by Hohenberg and 
Kohn \cite{Hohenberg}, the exact ground state energy of a system of $N$ electrons can be formally 
written as a functional $E[\rho]$ of only the electron density $\rho({\bf r})$, a function of three variables, 
and the external field $V_{ext}({\bf r})$ \cite{Parr,Dreizler}.
Kohn and Sham (KS) \cite{Kohn} partitioned $E[\rho]$ into the sum of three terms,
the classical Hartree energy $E_H[\rho]$ (readily expressed in terms of the density),
the kinetic energy density functional (KEDF) $T_{s}[\rho]$ of a model system of noninteracting
electrons at the same density, and the much smaller remainder,
the so-called exchange-correlation energy functional $E_{xc}[\rho]$.
KS showed that the numerical value of $T_{s}[\rho]$ could be determined
exactly, not directly from the density itself, but by using a set of $N$ one-electron
wave functions (orbitals) to solve the $N$ coupled KS
equations that describe the noninteracting system. By using these results along
with relatively simple approximations
for the small $E_{xc}[\rho]$ term, the resulting KS-DFT has proven
successful in many applications.

However, because of the use of the KS orbitals, KS-DFT typically scales as $O(N^{3})$, 
which limits the applicability of KS-DFT. By using localized orbitals some orbital-based linear-scaling
methods have been developed that are especially successful for insulating systems with 
large band gaps \cite{LinearScaling,Goedecker}. However, the prefactors of these
types of approaches are usually so large that their computational costs become cheaper than the 
traditional KS-DFT (with $O(N^{3})$ scaling) only for very large systems.

This problem could be circumvented if we could develop an accurate ``orbital-free"
density functional theory (OF-DFT), 
where the kinetic energy is expressed in terms of the electron density alone \cite{Parr,Dreizler,WatsonCarter,Kaxiras}.  
Of course, given its large magnitude, this is a very difficult task,
and indeed simple local approximations like that used in the Thomas-Fermi (TF) model \cite{Thomas,Thomas2}
have proved very inaccurate.

Through the recent efforts of many workers \cite{WT,WGC,WGC2,Zhou3,Herring,WC,CAT,GAC,GAC2,Madden,Perrot,Alonso,Alonso2},
we now have significantly better approximations for the KEDF.
There have been two main advances. The first is the use of nonlocal KEDFs that reproduce known exact
results both for very slowly varying or very rapidly varying fields, and that give the exact linear response (LR)
of the density of uniform model systems to small perturbations. The second is to focus on the
more slowly-varying valence electron density as described by weak local 
pseudopotentials \cite{LPS,LPS2,LPS3,Goodwin,LPS4,LPS5,LPS6,Zhou,Zhou2,BWang,ChaiWeeks2}. For OF-DFT with LR-based KEDFs, 
the use of such pseudopotentials not only can reduce the computational 
cost relative to all-electron calculations,
but also can improve its accuracy, since the valence system will be closer to the LR
regime where $T_{s}[\rho]$ is designed to be accurate \cite{Zhou,BWang}.
Very promising results using such OF-DFT methods
have been obtained for a variety of nearly free-electron-like metals.
  
However, a more accurate
treatment of the KEDF is still needed in other applications with significant variations in the density.
The main problem is that the exact
$T_{s}[\rho]$ is highly nonlocal, and we have little idea of the functional form
of the nonlocality for densities far from the linear response regime.

In earlier work, Chai and Weeks (CW) \cite{ChaiWeeks2,ChaiWeeks} proposed a different way to attack this 
basic problem. In the usual approach, approximate forms for $T_{s}[\rho]$ and $E_{xc}[\rho]$
are assumed, and the density $\rho({\bf r})$ is
obtained from the variational principle (Euler equation) associated with minimizing $E[\rho]$:  
\begin{equation}
\mu=V_{T_{s}}({\bf r};[\rho])+V_{eff}({\bf r};[\rho]).\label{eq:2}
\end{equation}
The total energy of the inhomogeneous system is then determined
from the energy functional $E[\rho]$. All other physical quantities
related to the ground-state density could also be computed. Here $\mu$
is the chemical potential (the Lagrange multiplier associated with
the normalization condition $\int\rho({\bf r})d{\bf r}=N$), and $V_{eff}({\bf r};[\rho])$
is an effective one-body potential defined by 
\begin{eqnarray}
V_{eff}({\bf r};[\rho]) & \equiv & V_{H}({\bf r};[\rho])+V_{xc}({\bf r};[\rho])+V_{ext}({\bf r}),\label{eq:4}
\end{eqnarray} 
where 
\begin{equation}
V_{H}({\bf r};[\rho])\equiv \int\frac{\rho({\bf r'})}{|{\bf r}-{\bf r'}|}d{\bf r'}\label{eq:4a}
\end{equation}
is the \emph{Hartree potential}, and $V_{xc}({\bf r};[\rho])\equiv\delta E_{xc}[\rho]/\delta\rho({\bf r})$
is the \emph{exchange-correlation potential}. 
Similarly we interpret
\begin{equation}
V_{T_{s}}({\bf r};[\rho])\equiv\delta T_{s}[\rho]/\delta\rho({\bf r})\label{eq:3}
\end{equation}
as the \emph{kinetic potential} (KP)\ arising from the KEDF \cite{KingHandy,KingHandy2}.

In principle this process could be reversed, and the exact $T_{s}[\rho]$ could be determined 
from the KP $V_{T_{s}}({\bf r};[\rho])$ by functional integration over density
changes in all regions of space \cite{ChaiWeeks2,Herring,KingHandy,KingHandy2}.
Because of this integration, $T_{s}[\rho]$ is a functional that depends
on the density everywhere even in the simplest case, exemplified by TF theory,
where $V_{T_{s}}({\bf r};[\rho])$ is a simple function of the local density at ${\bf r}$ only.

Of course, the true $V_{T_{s}}({\bf r};[\rho])$ itself is much more complicated and
must be a nonlocal functional of the density as well.
However it seems reasonable to assume that it depends most
strongly on the density in some local region
around ${\bf r}$. Recent detailed arguments \cite{Holas,Holas2} suggest
that the KP in this sense is more local than is the KEDF. Since most problems in
devising accurate approximations for $T_{s}[\rho]$ have arisen from
the nonlocality, this suggests that it could be worthwhile to try to develop
approximations for the more local KP $V_{T_{s}}({\bf r};[\rho])$.

To that end, following ideas first introduced for the OF-KEDF's,
CW proposed two different nonlocal KPs for atomic systems that reduce to known
exact forms for both slowly varying and rapidly varying perturbations and also
reproduce exact results for the linear response of the density of
the homogeneous system to small perturbations  \cite{ChaiWeeks2}.
The latter requirement introduces nonlocality into the resulting KPs.
CW also proposed two efficient pathways
to determine the numerical value of the kinetic energy by integration of the KPs,
as discussed below in Sec.~\ref{sec:pathway}.

Using the CW KPs and the most accurate (Herring) pathway
along with weak \textit{ab initio} local pseudopotentials for the 
valence component of the electron density, CW obtained results for the total energies 
and ionization energies of atoms, and for the shell structure in the atomic radial density profiles that are in 
better agreement with calculations using the full Kohn-Sham theory than earlier KEDFs were able to produce.
For example, the ionization energies of the first and second row atoms predicted by the CW KPs have
average errors of 1.8 eV and 2.1 eV for the two different KPs,
compared with the full KS-DFT, while the CAT KEDF \cite{GAC2}, a 
nonlocal KEDF, has an average error of 3.9 eV. Similarly, errors in the valence energies predicted 
by CAT KEDF are almost twice as large than those predicted by the CW KPs \cite{ChaiWeeks2}.  

In this paper, we use the CW KPs to study two very
different solid-state systems, a nearly-free-electron-like 
metal (solid Al) and a covalent material (solid Si). As discussed below,
the original forms of these KPs use the local Fermi wave vector. This
permits their use in atomic and molecular systems where the electron density
vanishes far from from the nuclei. However, this also leads to
a quadratic scaling with system size \cite{ChaiWeeks2}, which limits 
their usefulness for very large extended systems. Following earlier work on the KEDF \cite{WGC2,Zhou3},
we introduce in Sec.\ \ref{sec:linear} an expansion method to reduce the computational 
cost of evaluating these nonlocal KPs, and the truncated expansions essentially scale
linearly with system size. The density pathway by which the kinetic energy is obtained from the
KP is described in Sec.\ \ref{sec:pathway}. Section \ref{sec:results} compares
the results of this method with the full KS-DFT
and with other approaches using KEDFs and gives our conclusions. 

\section{Linear Scaling Methods}
\label{sec:linear}
CW introduced two different nonlocal KPs \cite{ChaiWeeks2}, one denoted LQ because
it satisfies known exact results to second order at low wavevectors (``Low $q$''), and the other
HQ with similar behavior at high wavevectors (``High $q$''). The nonlocality is generated
by the requirement that these KPs also reproduce the exact linear response function of a uniform
electron gas at the the local Fermi wave vector (LFWV) $k_{F}({\bf r})$, defined in terms of the
electron density by
\begin{equation}
k_{F}({\bf r})\equiv(3\pi^{2}\rho({\bf r}))^{1/3}.
\label{eq:lfwv}
\end{equation}
Both potentials can be compactly described using the following
generalized form with different values for a parameter $\alpha$:
\begin{equation}
\begin{array}{ll}
V_{\alpha}({\bf r};[\rho],k_{F}({\bf r}))=V_{TF}({\bf r};[\rho])+V_{W}({\bf r};[\rho])\textbf{\medskip}\\
\;\;+\frac{10}{9\alpha}C_{F}\rho^{\beta}({\bf r}){\displaystyle \int f(|{\bf r}-{\bf r'}|;k_{F}({\bf r}))\,\rho^{\alpha}({\bf r'})d{\bf r'}}
\label{eq:hlq}
\end{array}
\end{equation}
Here $\beta \equiv (2/3-\alpha)$. The HQ model uses $\alpha = 2/3$ and the LQ model has $\alpha = 1/2$.
$V_{TF}({\bf r};[\rho])$ and $V_{W}({\bf r};[\rho])$ are the TF KP and the 
von Weizs\"{a}cker (W) KP \cite{Weizsacker} respectively, and $C_{F}$
is a numerical coefficient associated with the TF KEDF. See Ref.\ \cite{ChaiWeeks2}
for a detailed discussion.

The weight function $f(|{\bf r}-{\bf r'}|;k_{F}({\bf r}))$ is defined by the following inverse
Fourier-transform-like integral:
\begin{equation}
f(|{\bf r}-{\bf r'}|;k_{F}({\bf r})) \equiv \frac{1}{(2\pi)^{3}} \int\hat{f}(k/2k_{F}({\bf r}))  e^{-i{\bf k}\cdot({\bf r}-{\bf r'})}d{\bf k}.
\label{eq:deff}
\end{equation}
Here
\begin{equation}
\hat{f}(q)=F_{L}(q)-3q^{2}-1.
\label{eq:weightfunction}
\end{equation}
is directly related
to the inverse linear response function $F_{L}(q)$ of the uniform electron gas
with density $\rho_0$ at reduced wave vector $q \equiv k/2k_{F0}$, with
$k_{F0} \equiv (3\pi^{2}\rho_0)^{1/3}$ \cite{Lindhard}:
\begin{equation}
F_{L}(q)\equiv\left[\frac{1}{2}+\frac{1-q^{2}}{4q}\ln\left|\frac{1+q}{1-q}\right|\right]^{-1}.
\label{eq:12a}
\end{equation}

Clearly $k_{F0}$ is replaced by the LFWV in the the definition of the weight
function $f(|{\bf r}-{\bf r'}|;k_{F}({\bf r}))$ in Eq.\ (\ref{eq:deff}). Given this definition it is
straightforward to rewrite the last term in Eq.\ (\ref{eq:hlq}) in Fourier space as
\begin{equation}
\frac{10}{9 \alpha (2\pi)^{3}}C_{F}\rho^{\beta}({\bf r})\int\hat{f}(k/2k_{F}({\bf r}))\rho^{\alpha}({\bf k})e^{-i{\bf k}\cdot{\bf r}}d{\bf k}.
\label{eq:hlqi}
\end{equation}
Because of the $k_{F}({\bf r})$ term in Eq.\ (\ref{eq:hlqi}), the integration must
be done on a grid over a range of $\bf r$, and all these models 
scale quadratically. 

In order to achieve the desirable linear scaling for fast computation in large solid-state systems, we 
follow the work of Wang \textit{et al.} \cite{WGC2} and apply a Taylor series expansion to the weight functional
$f(|{\bf r}-{\bf r'}|;k_{F}({\bf r}))$ with respect to a reference
density $\rho_{*}$. For extended systems, where $\rho({\bf r})$ is not significantly
different from the average density $\rho_{0}$, the natural choice
of $\rho_{*}$ is $\rho_{0}$. Throughout this paper, we
choose $\rho_{*}=\rho_{0}$ in all of our calculations. 

The weight function $f(|{\bf r}-{\bf r'}|;k_{F}({\bf r}))$  can be expanded with respect to $\rho_{*}$ as:
\begin{equation}
\begin{array}{ll}
f(|{\bf r}-{\bf r'}|;k_{F}({\bf r}))=f(|{\bf r}-{\bf r'}|;k_{F}^{*}) \textbf{\medskip}\\
\;\;\;\;\;\;\;\;\;\;\;\;\;\;+f^{(1)}(|{\bf r}-{\bf r'}|;k_{F}^{*})(\rho({\bf r})-\rho_{*})\textbf{\medskip}\\
\;\;\;\;\;\;\;\;\;\;\;\;\;\;\;\;+\frac{1}{2}f^{(2)}(|{\bf r}-{\bf r'}|;k_{F}^{*})(\rho({\bf r})-\rho_{*})^{2}+\cdots
\end{array}
\label{eq:4.5.1}
\end{equation}
where $k_{F}^{*}\equiv(3\pi^{2}\rho_{*})^{1/3}$, and $f^{(n)}(|{\bf r}-{\bf r'}|;k_{F}^{*})=\partial^{n}f(|{\bf r}-{\bf r'}|;k_{F}({\bf r}))/\partial\rho^{n}({\bf r})\mid_{\rho_{*}}$
is the $n$th order derivative of $f(|{\bf r}-{\bf r'}|;k_{F}({\bf r}))$
with respect to $\rho({\bf r})$, and is evaluated at $\rho_{*}$.
Their functional forms, up to second order, in reciprocal space are
\begin{equation}
\hat{f}^{(1)}(q_{*})=-\frac{q_{*}}{3\rho_{*}}\hat{f}^{\prime}(q_{*})\label{eq:4.5.2}\end{equation}
and
\begin{equation}
\hat{f}^{(2)}(q_{*})=\frac{q_{*}^{2}\hat{f}^{\prime\prime}(q_{*})+4q_{*}\hat{f}^{\prime}(q_{*})}{(3\rho_{*})^{2}}\label{eq:4.5.3}\end{equation}
where $q_{*}={k}/(2k_{F}^{*})$, and $\hat{f}^{\prime}(q_{*})$ and $\hat{f}^{\prime\prime}(q_{*})$
are the first and the second derivative of $\hat{f}(q)$ with respect to $q_{*}$.
Since the analytical form of $\hat{f}(q)$
is available in Eq.\ (\ref{eq:weightfunction}), all the terms
needed in the Taylor series expansion $\hat{f}^{(n)}(q)$ can be obtained analytically.
This simplicity is one of the main advantages of the KP method over related methods
that use the KEDF, where these terms have usually been be obtained numerically, with one
recent exception \cite{analytic}.

By carrying out the Taylor series expansion to $n$th order in
Eq.\ (\ref{eq:4.5.1}), and inserting it in Eq.\ (\ref{eq:hlq}),
one can then take out the $(\rho({\bf r})-\rho_{*})^{m}$ factor of the $m$th
term from its integral (where $m$ is a nonnegative integer, and $m\leq n$),
and the remaining integral becomes:
\begin{equation}
\int f^{(m)}(|{\bf r}-{\bf r'}|;k_{F}^{*})\,\,\rho^{\alpha}({\bf r'})d{\bf r'}\label{eq:4.5.5}\end{equation}
The integral in Eq.\ (\ref{eq:4.5.5}) can be easily evaluated by a
simple fast Fourier transform (FFT). Therefore, to compute the nonlocal term in Eq.\ (\ref{eq:hlq})
using this scheme, one needs to evaluate a total of $n+2$ FFT's, including
the FFT of $\rho^{\alpha}$. Since all of the terms can be computed
by FFT's, this scheme essentially scales linearly $O(M\ln M)$ with
system size, where $M$ is the number of grid points.

The corresponding linear-scaling HQ ($\alpha = 2/3$) and LQ ($\alpha = 1/2$) KPs
with $\rho_{*}=\rho_{0}$ can be written as:
\begin{equation}
\begin{array}{ll}
V_{\alpha }^{lin}({\bf r};[\rho ],k_{F0})=V_{TF}({\bf r};[\rho ])+V_{W}({\bf r%
};[\rho ]){\bf \medskip } &  \\ 
\;\;+\frac{10}{9\alpha }C_{F}\rho ^{\beta}({\bf r})\displaystyle %
\left\{ {\int f(|{\bf r}-{\bf r^{\prime }}|;k_{F0})\,\rho ^{\alpha }({\bf %
r^{\prime }})d{\bf r^{\prime }}}\right. {\bf \medskip } &  \\ 
\;\;\;\;+(\rho ({\bf r})-\rho _{0}){\displaystyle \int f^{(1)}(|{\bf r}-{\bf %
r^{\prime }}|;k_{F0})\,\,\rho ^{\alpha }({\bf r^{\prime }})d{\bf r^{\prime }}}%
{\bf \medskip } &  \\ 
\;\;\;\;\;\;+\left. \frac{1}{2}(\rho ({\bf r})-\rho _{0})^{2}{\displaystyle %
\int f^{(2)}(|{\bf r}-{\bf r^{\prime }}|;k_{F0})\,\,\rho ^{\alpha }({\bf %
r^{\prime }})d{\bf r^{\prime }}+\cdots }\right\}  & 
\end{array}
\label{eq:hlqlinear}
\end{equation}
where $k_{F0}\equiv(3\pi^{2}\rho_{0})^{1/3}$ is the uniform Fermi wave vector.
For simple nearly-free-electron metals like Al, the expansion method works very well,
and indeed often only the zeroth order term is needed.
For extended systems with
large density variations over space, this expansion method could experience convergence
problems similar to those that have been seen by other workers \cite{WGC2,WC,Zhou,Zhou2,Zhou3}.

However, in those cases, both the use of the LFWV in Eq.\ \ref{eq:lfwv}
and the basic linear response treatment of the nonlocality are probably inadequate
as well.
A two-body Fermi wavevector $k_{F}({\bf r},{\bf r'})$ like the one used in
the WGC KEDF \cite{WGC2} could be introduced, but
this functional form represents an additional approximation that
does not systematically improve the underlying linear response treatment of the nonlocality.
As we will see, the treatment of extended systems with large density variations
remains a major challenge for all OF-DFT methods.

\section{Density Pathway}
\label{sec:pathway}
As discussed in Ref.\ \cite{ChaiWeeks2}, an integration pathway is
needed to determine the value of $T_{s}[\rho]$
from a given $V_{T_{s}}({\bf r};[\rho])$ \cite{Pratt,Pratt2,Pratt3,Chen}.
If the exact $V_{T_{s}}({\bf r};[\rho])$ is used and the integration is
carried out exactly, then all pathways would give the 
same exact result for $T_{s}[\rho]$. An approximate KP can give the same (approximate)
value for the kinetic energy independent of pathway only if it arises from
functionally differentiating a KEDF (the usual OFDFT approach) or
equivalently if it exactly satisfies the nonlocal
consistency conditions given by Herring \cite{Herring} in his Eq.\ (24). As he points out, 
it is not easy to satisfy these formal conditions for general functionals,
and we do not try to do so here.

Instead, we generate nonlocality by imposing a more easily implemented
and physically suggestive condition, requiring that linear response theory is exactly
satisfied for small perturbations about a uniform system.
Thus our results will be independent of pathway in the linear regime but
the kinetic energy obtained from
integration of our approximate KPs for systems with large density variations will
in general depend on the particular integration pathway used,
as discussed in detail in Ref.\ \cite{ChaiWeeks2}. Experience with the classical analogue of the
KP method applied to nonuniform hard sphere fluids \cite{Chen} has shown that errors
from this path dependence can be small (e.g., less than one percent for the surface tension at a
hard wall) when used with particular pathways that do not excessively weight
regions poorly described by an approximate theory for the nonuniform density.

Two efficient pathways to obtain the kinetic energy from an approximate
KP for atomic and molecular systems
were discussed in Ref.\ \cite{ChaiWeeks2}.
The simplest and most accurate pathway, due to Herring \cite{Herring},
automatically satisfies the virial theorem.
However it does not apply to extended systems in its original form and
we have not yet found the appropriate generalization.
Here we use the alternate linear density
pathway, where there is a linear scaling of the electron density
\begin{equation}
\rho_{\lambda}({\bf r})\equiv \rho_{0}+ \lambda [ \rho({\bf r}) - \rho_{0}].
\label{eq:lindensity}
\end{equation}
If the exact KP were used, both pathways would give the exact kinetic energy.

By inserting the HQ and LQ KPs of Eq.\ (\ref{eq:hlqlinear})
into the linear density pathway (see Eqs.\ (9)-(11) in Ref.\ \cite{ChaiWeeks2}), 
the kinetic energy can be computed as
\begin{equation}
\begin{array}{ll}
T_{\alpha}=T_{\lambda=0}+\int_{0}^{1}d\lambda\int d{\bf r}[\rho({\bf r})-\rho_{0}] V_{\alpha}^{lin}({\bf r};[\rho_{\lambda}],k_{F0}).
\end{array}
\label{eq:4.6.2.2}
\end{equation}
Here $T_{\lambda=0}$ is the kinetic energy for uniform system, i.e., the Thomas-Fermi kinetic energy $T_{TF}[\rho_{0}]$.
Since the $V_{TF}({\bf r};[\rho_{\lambda}])$ and $V_{W}({\bf r};[\rho_{\lambda}])$
in Eq.\ (\ref{eq:4.6.2.2}) arise from the functional derivatives of
the known $T_{TF}[\rho]$ and $T_{W}[\rho]$ functionals respectively, these terms can
be integrated exactly and lead to the TF and W KEDFs.

Equation (\ref{eq:4.6.2.2}) then becomes
\begin{equation}
\begin{array}{l}
T_{\alpha }[\rho ]= T_{TF}[\rho ]+T_{W}[\rho ]+%
\displaystyle \int %
_{0}^{1}d\lambda \int d{\bf r}[\rho ({\bf r})-\rho _{0}]
\\ 
\;\;\;\times \frac{10}{9\alpha }C_{F}\rho^{\beta} _{\lambda }({\bf r})\left\{ 
\displaystyle \int %
d{\bf r^{\prime }}f(|{\bf r}-{\bf r^{\prime }}|;k_{F0})\rho^{\alpha } _{\lambda }({\bf %
r^{\prime }})\right.  \\ 
\;\;\;+(\rho _{\lambda }({\bf r})-\rho _{0})%
\displaystyle \int %
d{\bf r^{\prime }}f^{(1)}(|{\bf r}-{\bf r^{\prime }}|;k_{F0})\rho^{\alpha } _{\lambda
}({\bf r^{\prime }})\\ 
\;\;\;+\left. \frac{1}{2}(\rho _{\lambda }({\bf r})-\rho _{0})^{2}%
\displaystyle \int %
d{\bf r^{\prime }}f^{(2)}(|{\bf r}-{\bf r^{\prime }}|;k_{F0})\rho^{\alpha } _{\lambda
}({\bf r^{\prime }})+\cdots \right\} 
\end{array}
\label{eq:Ehlqlinear}
\end{equation}
where $T_{\lambda=0}=T_{TF}[\rho_{0}]$ term
is absorbed in the $T_{TF}[\rho]$ term. We use Eq.\ (\ref{eq:Ehlqlinear}) to evaluate the kinetic energy of
our linear-scaling HQ ($\alpha = 2/3$) and LQ ($\alpha = 1/2$)  KPs in all the calculations here.

\section{Results}
\label{sec:results}
In solids, the external potential $V_{ext}({\bf r})$ can be regarded as a linear
combination of the special array of local atomic pseudopotentials
centered at each ion position ${\bf R}_{I}$.
Different arrays of ${\bf R}_{I}$ lead to different phases, such
as face-centered cubic (fcc), diamond (dia), body-centered cubic
(bcc), simple cubic (sc), and so on. We calculate the binding energies of Al and Si at these four different
phases, and compare our results with other KEDFs and KS-DFT.
We use periodic boundary conditions with a cubic supercell containing 4 atoms for fcc, 8 atoms for dia,
2 atoms for bcc, and 8 atoms for sc. All calculations are spin-restricted and use the local 
density approximation (LDA) \cite{Dirac,Ceperley,Ceperley2,Perdew} for the exchange-correlation functional.

\begin{table} [t]

\caption{\label{table:5.6} Lattice parameters (\AA) for
bulk Al. The results for the KS, WT, and WGC models are taken from Ref.\ \cite{WGC2}.}

\begin{tabular}{cccccc}
\hline
\hline 
\ \ Al \ \ &
\ \ KS \ \ &
\ \ LQ \ \ & 
\ \ HQ \ \ &
\ \ WT \ \ &
\ \ WGC \ \ \tabularnewline
\hline
fcc&
$4.03$&
$4.04$&
$4.04$& 
$4.04$& 
$4.03$\tabularnewline
bcc&
$3.23$&
$3.23$&
$3.23$&
$3.23$&
$3.22$\tabularnewline
sc&
$5.33$&
$5.31$&
$5.36$&
$5.33$&
$5.38$\tabularnewline
dia&
$5.84$&
$5.91$&
$5.90$&
$5.94$&
$5.92$\tabularnewline
\hline
\hline
\end{tabular}
\end{table}

\begin{table}

\caption{\label{table:5.6a} Energy per atom (eV) for bulk Al. The first
row is the energy for the fcc structure, while other rows are energy
difference from the fcc structure. The results for the KS, WT, and WGC models are taken from Ref.\ \cite{WGC2}.}

\begin{tabular}{cccccc}
\hline
\hline
Al&
KS&
LQ&
HQ&
WT&
WGC\tabularnewline
\hline
fcc&
$-58.336$&
$-58.303$&
$-58.314$&
$-58.331$&
$-58.331$\tabularnewline
bcc&
$0.068$&
$0.053$&
$0.057$&
$0.060$&
$0.066$\tabularnewline
sc&
$0.250$&
$0.253$&
$0.253$&
$0.227$&
$0.217$\tabularnewline
dia&
$0.599$&
$0.712$&
$0.751$&
$0.673$&
$0.584$\tabularnewline
\hline
\hline
\end{tabular}
\end{table}

For bulk Al, the empirical Goodwin-Needs-Heine (GNH) local pseudopotential
\cite{Goodwin} is used, and a plane wave kinetic energy cutoff of
600 eV is used to converge the electron density. Here it is sufficient to use only the zeroth-order
linear-scaling HQ and LQ KPs (see Eq.(\ref {eq:hlqlinear}))
in the calculations, as verified by comparison with results of a first order calculation.
Our results are compared with the WT \cite{WT} and WGC \cite{WGC2} KEDFs, and KS-DFT,
which were previously computed \cite{WGC2}, using the same
local pseudopotential. (The atomic-based local pseudopotentials in Ref.\ \cite{ChaiWeeks2} 
could also be used and gives very similar results.)  As can be seen in Tables \ref{table:5.6}
and \ref{table:5.6a}, all the LR-based models perform similarly,
and agree well with KS-DFT. The phase ordering is correct, and
aside from the high energy diamond phase, the lattice parameters are close to the KS results.

\begin{table} [tbp]
\caption{\label{table:5.6b} Lattice parameters (\AA) for
bulk Si. The results for the KS and WGC models are taken from Ref.\ \cite{Zhou}, 
and those for the WGC2 model are taken from Ref.\ \cite{Zhou3}.}

\begin{tabular}{cccccc}
\hline
\hline
\ \ Si \ \ &
\ \ KS \ \ &
\ \ LQ \ \ &
\ \ HQ \ \ &
\ \ WGC \ \ &
\ \ WGC2 \ \ \tabularnewline
\hline
dia&
$5.38$&
$5.27$&
$5.29$&
$5.77$&
$5.57$\tabularnewline
bcc&
$3.29$&
$3.06$&
$3.06$&
$3.29$&
$3.32$\tabularnewline
sc&
$4.99$&
$4.98$&
$4.98$&
$5.01$&
$5.06$\tabularnewline
fcc&
$3.83$&
$3.83$&
$3.82$&
$3.80$&
$3.80$\tabularnewline
\hline
\hline
\end{tabular}
\end{table}

\begin{table} [!tbp]
\caption{\label{table:5.6c} Energy per atom (eV) for bulk Si. The first
row is the energy for the dia structure, while other rows are energy
difference from the dia structure. The results for the KS and WGC models are taken from Ref.\ \cite{Zhou},
and those for the WGC2 model are taken from Ref.\ \cite{Zhou3}.}

\begin{tabular}{cccccc}
\hline
\hline
Si&
KS&
LQ&
HQ&
WGC&
WGC2\tabularnewline
\hline
dia&
$-110.234$&
$-109.167$&
$-109.282$&
$-110.345$&
$-110.220$\tabularnewline
bcc&
$ 0.165$&
$-0.553$&
$-0.437$&
$ 0.537$&
$ 0.267$\tabularnewline
sc&
$ 0.303$&
$-0.586$&
$-0.531$&
$ 0.506$&
$ 0.308$\tabularnewline
fcc&
$ 0.457$&
$-0.584$&
$-0.478$&
$ 0.571$&
$ 0.444$\tabularnewline
\hline
\hline
\end{tabular}
\end{table}

To assess the performance of the HQ and LQ KPs in covalent systems like solid Si,
we use the bulk local pseudopotential (BLPS) developed by
Zhou \textit{et al.} \cite{Zhou} together with a plane wave kinetic energy cutoff
of 2000 eV for converging the electron density. We used the first-order
 linear-scaling HQ and LQ KPs in the Si calculations.
Our results are compared first with the original WGC
KEDF \cite{WGC2}, and with the KS-DFT,
which were previously computed
\cite{Zhou}, using the same local pseudopotential. As can
be seen in Tables \ref{table:5.6b} and \ref{table:5.6c}, all these LR-based
models perform worse than for Al, when compared with KS-DFT.
The phase ordering is incorrect, and the lattice parameters of the
four phases only qualitatively match with the KS results, although the
WGC KEDF at least obtains the correct diamond ground state.

Because we use approximate KPs, our estimate for $T_{s}[\rho]$ will depend on the integration
pathway, and it is important to try to find a particular pathway that is relatively insensitive to 
the errors that exist in our KPs. For atomic systems, we were able to use the Herring pathway, which automatically satisfies the virial theorem,
and good results were found \cite{ChaiWeeks2}. As noted above, a generalization of the Herring pathway for solids is not
yet available, and in this paper we used the linear density pathway. While this pathway is numerically efficient, and gives exact
results when the exact KP is used, we expect less accurate results when used with our approximate KPs
since the virial theorem is not satisfied.

Indeed this is the case for atomic systems, where results from both pathways can be compared.
In Tables \ref{table:path}, we show the pathway dependence of the full LQ KP
(with \textit{ab initio} local pseudopotentials \cite{ChaiWeeks2}) 
in atomic systems. The kinetic energies of the three atoms (Al, Si and Ar),
using the full LQ KP, are evaluated by the Herring and the density pathways,
and are compared with the KS results. Clearly, the density pathway gives
significantly worse kinetic energy than the Herring's pathway, 
and its errors increase when the densities of the systems are more rapidly varying.
This suggests that the large deviations of the LQ and HQ results for solid Si (dia) in Tables \ref{table:5.6c} may be partially due to the
use of this pathway. This pathway dependence also seems likely to influence the phase orderings. Thus the generalization
of the accurate Herring pathway to solids seems a promising focus for future research.

\begin{table} [!t]
\caption{\label{table:path} The kinetic energy T$_{s}$[$\rho$] (eV) of the three atoms (Al, Si, and Ar) in \textit{ab initio} local 
pseudopotential calculations \cite{ChaiWeeks2}, using the KS method and the full LQ model (evaluated by both the Herring's 
and the density pathways). 
MAE, the mean absolute errors (relative to the KS method) of the full LQ model computed by the two pathways are given 
at the bottom of their respective columns.}

\begin{tabular}{lrrr}
\hline
\hline
&
\  \ KS&
\  \ LQ$_{H}$  &
\  \  LQ$_{den}$   \tabularnewline
\hline
Al \  \ &
$21.164$&
$21.823$&
$17.563$\tabularnewline
Si \  \ &
$40.985$&
$41.264$&
$33.874$\tabularnewline
Ar \  \ &
$225.609$&
$221.035$&
$196.495$\tabularnewline
MAE \  \ &
$$&
$1.838$&
$13.276$\tabularnewline
\hline
\hline
\end{tabular}
\end{table}

Since Si is a covalent material, the density
inside the covalent bond regions is quite different from that outside. Even with a local
pseudopotential this system may be outside the linear response regime,
or the results at least may depend sensitively of the choice of the one- or two-body FWV.
In particular, recent work \cite{Zhou3} has shown that much improved results
for the phase energies can be obtained for bulk Si using
the WGC KEDF with a FWV mixing parameter and reference 
density optimized for the covalent phases of Si, rather than
the original FWV mixing parameter optimized for Al \cite{WGC2}.
Results with this new parameter set, denoted here WGC2, are also given
in Tables \ref{table:5.6b} and \ref{table:5.6c}, where we see that the phase orderings
and energy differences for Si phases using WGC2 are in very good agreement with KS-DFT.
However, ultimately it is desirable to find a general
KEDF or KP that does not depend on special
properties of the system, particularly when considering more complicated
systems with defects or surfaces.
In our opinion, developing a truly universal and yet accurate KEDF or KP 
for covalent materials remains an outstanding problem.

\section{Conclusion}

In summary, CW have previously demonstrated that the nonlocal HQ and LQ KPs 
work well for isolated atoms and ions \cite{ChaiWeeks2}, and now we show they can be used with
no additional parameterization in solid-state systems. Very good results are found for Al.
However, problems of inaccurate phase ordering and bond lengths in
Si are found, and a better treatment of nonlocality beyond the linear
response regime may be needed for covalent systems with significant density variations.
The simpler and more local form of the OF KP could be useful in future developments.
If it is possible to develop a new
pathway to obtain the kinetic energy in extended systems that satisfies the virial theorem,
analogous to the accurate Herring pathway for atomic systems  \cite{ChaiWeeks2},
we believe even the present LR-based KP's would likely give more accurate phase
energies. Further work along these lines is called for.

\begin{acknowledgments}
This work at the University of Maryland has been supported by the NSF Grants No. CHE01-11104 and  CHE05-17818, and by
the NSF-MRSEC under Grants No. DMR 00-80008 and DMR05-20471. E.A.C. is grateful for support from the NSF under Grant No. CHE05-17359 for this work.
One of the authors (J.D.C.) acknowledges the support from the UMCP Graduate School program,
the IPST Alexander program, and the CHPH Block Grant Supplemental program.
\end{acknowledgments}

\bibliographystyle{jcp}

\begin{thebibliography}{10}

\bibitem{Hohenberg}  P. Hohenberg and W. Kohn, Phys. Rev. {\textbf{136}}, B864 (1964).

\bibitem{Parr}  R. G. Parr and W. Yang, {\emph{Density-Functional Theory of Atoms and Molecules}}, (Oxford University Press, 1989).

\bibitem{Dreizler}  R. M. Dreizler and E. K. U. Gross, {\emph{Density Functional Theory: An Approach to the Quantum Many Body Problem}},
(Springer-Verlag, Berlin, 1990).

\bibitem{Kohn}  W. Kohn and L. J. Sham, Phys. Rev. {\textbf{140}}, A1133 (1965).

\bibitem{LinearScaling} W. Yang, Phys. Rev. Lett. {\textbf{66}}, 1438 (1991).

\bibitem{Goedecker} For a recent review, see S. Goedecker, Rev. Mod. Phys. {\textbf{71}}, 1085 (1999).

\bibitem{WatsonCarter} S. C. Watson and E. A. Carter, Comput. Phys. Commun. {\textbf{128}}, 67 (2000).

\bibitem{Kaxiras}  N. Choly and E. Kaxiras, Solid State Commun. {\textbf{121}}, 281 (2002).

\bibitem{Thomas}  L. H. Thomas, Proc. Cambridge Phil. Soc. {\textbf{23}}, 542 (1927).

\bibitem{Thomas2} E. Fermi, Z. Phys. {\textbf{48}}, 73 (1928).

\bibitem{WGC}  Y. A. Wang, N. Govind, and E. A. Carter, Phys. Rev. {\textbf{B 58}}, 13465 (1998); 
ibid. {\textbf{60}}, 17162(E) (1999); ibid. {\textbf{64}}, 129901(E) (2001).

\bibitem{WGC2}  Y. A. Wang, N. Govind, and E. A. Carter, Phys. Rev. {\textbf{B 60}}, 16350 (1999);
ibid. {\textbf{64}}, 089903(E) (2001).

\bibitem{WC}  See e.g., Y. A. Wang and E. A. Carter, in {\emph{Theoretical Methods in Condensed Phase Chemistry}}, edited by S. D. Schwartz,
{\emph{Progress in Theoretical Chemistry and Physics}}, (Kluwer, Boston, 2000) p. 117, and references therein.

\bibitem{Zhou3}  B. Zhou, V. L. Ligneres, and E. A. Carter, J. Chem. Phys. {\textbf{122}}, 044103 (2005).

\bibitem{CAT}  E. Chac\'on, J. E. Alvarellos, and P. Tarazona, Phys. Rev. {\textbf{B 32}}, 7868 (1985).

\bibitem{GAC}  P. Garc\'ia-Gonz\'alez, J. E. Alvarellos, and E. Chac\'on, Phys. Rev. {\textbf{A 54}}, 1897 (1996).

\bibitem{GAC2}  P. Garc\'ia-Gonz\'alez, J. E. Alvarellos, and E. Chac\'on, Phys. Rev. {\textbf{B 57}}, 4857 (1998).

\bibitem{Madden}  M. Foley and P. A. Madden, Phys. Rev. {\textbf{B 53}}, 10589 (1996).

\bibitem{Perrot}  F. Perrot, J. Phys.: Condens. Matter {\textbf{6}}, 431 (1994).

\bibitem{Alonso}  J. A. Alonso and L. A. Girifalco, Phys. Rev. {\textbf{B 17}}, 3735 (1978). 

\bibitem{Alonso2}M. D. Glossman, L. C. Balb\'as, and J. A. Alonso, Chem. Phys. {\textbf{196}}, 455 (1995).

\bibitem{Herring}  C. Herring, Phys. Rev. {\textbf{A 34}}, 2614 (1986).

\bibitem{WT}  L.-W. Wang and M. P. Teter, Phys. Rev. {\textbf{B 45}}, 13196 (1992).

\bibitem{Zhou}  B. Zhou, Y.A. Wang, and E. A. Carter, Phys. Rev. {\textbf{B 69}}, 125109 (2004).

\bibitem{Zhou2}  B. Zhou and E. A. Carter, J. Chem. Phys. {\textbf{122}}, 184108 (2005).

\bibitem{LPS}  W. C. Topp and J. J. Hopfield, Phys. Rev. {\textbf{B 7}}, 1295 (1973).

\bibitem{LPS2} J. A. Appelbaum and D. R. Hamann, Phys. Rev. {\textbf{8}}, 1777 (1973).

\bibitem{LPS3} M. Schl\"uter, J. R. Chelikowsky, S. G. Louie, and M. L. Cohen, Phys. Rev. {\textbf{12}}, 4200 (1975).

\bibitem{Goodwin} L. Goodwin, R. J. Needs, and V. Heine, J. Phys.: Condes. Matter \textbf{2}, 351 (1990).

\bibitem{LPS4} S. Watson, B. J. Jesson, E. A. Carter, and P. A. Madden, Europhys. Lett.  {\textbf{41}}, 37 (1998). 

\bibitem{LPS5} J. A. Anta and P. A. Madden, J. Phys.: Condens. Matter  {\textbf{11}}, 6099 (1999).

\bibitem{LPS6} D. J. Gonz\'alez, L. E. Gonz\'alez, J. M. L\'opez, and M. J. Stott, Phys. Rev.  {\textbf{B 65}}, 184201 (2002).

\bibitem{BWang}  B. Wang and M. J. Stott, Phys. Rev. {\textbf{B 68}}, 195102 (2003).

\bibitem{ChaiWeeks2}  J.-D. Chai and J.D. Weeks, Phys. Rev. {\textbf{B 75}}, 205122 (2007).

\bibitem{ChaiWeeks}  J.-D. Chai and J.D. Weeks, J. Phys. Chem {\textbf{B 108}}, 6870 (2004).

\bibitem{KingHandy}  R. A. King and N. C. Handy, Phys. Chem. Chem. Phys. {\textbf{2}}, 5049 (2000). 

\bibitem{KingHandy2} R. A. King and N. C. Handy, Mol. Phys. {\textbf{99}}, 1005 (2001).

\bibitem{Holas}   A. Holas and N. H. March, Phys. Rev. {\textbf{A 66}}, 066501 (2002).

\bibitem{Holas2} I. Lindgren and S. Salomonson, Phys. Rev. {\textbf{A 67}}, 056501 (2003).

\bibitem{Weizsacker}  C. F. von Weizs\"acker, Z. Physik {\textbf{96}}, 431 (1935).

\bibitem{Lindhard}  J. Lindhard, K. Dan. Vidensk. Selsk. Mat. Fys. Medd. {\textbf{28}}, 8 (1954).

\bibitem{analytic} G. Ho, V. L. Ligneres, and E. A. Carter, Phys. Rev. {\textbf{B 78}}, 045105 (2008).

\bibitem{Pratt}   L. R. Pratt, G. G. Hoffman, and R. A. Harris, J. Chem. Phys. {\textbf{88}}, 1818 (1988). 

\bibitem{Pratt2} L. R. Pratt, G. G. Hoffman, and R. A. Harris, J. Chem. Phys. {\textbf{92}}, 6687 (1990). 

\bibitem{Pratt3} G. G. Hoffman and L. R. Pratt, Mol. Phys. {\textbf{82}}, 245 (1994).

\bibitem{Chen}  Y.-G. Chen and J. D. Weeks, J. Chem. Phys. {\textbf{118}}, 7944 (2003).

\bibitem{Dirac}  P. A. M. Dirac, Proc. Cambridge Phil. Soc. {\textbf{26}}, 376 (1930).

\bibitem{Ceperley}  D. M. Ceperley, Phys. Rev. {\textbf{B 18}}, 3126 (1978). 

\bibitem{Ceperley2}D. M. Ceperley and B.J. Alder, Phys. Rev. Lett. {\textbf{45}} , 566 (1980).

\bibitem{Perdew}  J. P. Perdew and A. Zunger, Phys. Rev. {\textbf{B 23}}, 5048 (1981).

\end{thebibliography}

\end{document}